\documentclass[runningheads]{llncs}

\makeatletter
\def\footnoterule{\relax%
  \kern-5pt
  \hbox to \columnwidth{\hfill\vrule width 0.8\columnwidth height 0.4pt\hfill}
  \kern4.6pt}
\makeatother

\usepackage{graphicx}

\usepackage[utf8]{inputenc}

\usepackage{multirow}
\usepackage[utf8]{inputenc}
\usepackage{lipsum}
\usepackage{multicol}
\usepackage{graphicx}

\usepackage{listings}
\usepackage{xcolor}
\usepackage{hyperref}
\let\svthefootnote\thefootnote
\newcommand\freefootnote[1]{%
  \let\thefootnote\relax%
  \footnotetext{#1}%
  \let\thefootnote\svthefootnote%
}
\usepackage{indentfirst}

\begin{document}

\title{A Hardware Co-design Workflow for Scientific Instruments at the Edge}
\author{Kazutomo Yoshii,\inst{1} Rajesh Sankaran,\inst{1} Sebastian Strempfer,\inst{1} Maksim Levental,\inst{2} Mike Hammer,\inst{1} Antonino Miceli\inst{1}}
\institute{
Argonne National Laboratory, Lemont, IL\\
\and
University of Chicago, Chicago, IL}

\maketitle

\section{Abstract}
%
%
As spatial and temporal resolutions of scientific instruments  improve,  the explosion in the volume of data produced is becoming a key challenge. It can be a critical bottleneck for integration between scientific instruments at the edge and high-performance  computers/emerging accelerators. Placing data compression or reduction logic close  to the data source is a possible approach to solve the bottleneck.
However, the realization of such a solution requires the development of custom ASIC designs, which is still challenging in practice and tends to produce one-off implementations unusable beyond the initial intended scope. Therefore, as a feasibility study, we have been investigating a design workflow that allows us to explore algorithmically complex hardware designs and develop reusable hardware libraries for the needs of scientific instruments at the edge. Our vision is to cultivate our hardware development capability for streaming/dataflow hardware components that can be placed close to the data source to enable extreme data-intensive scientific experiments or environmental sensing. Furthermore, reducing data movement is essential to improving computing performance in general. Therefore, our co-design efforts on streaming hardware components can benefit computing applications other than scientific instruments. This vision paper discusses hardware specialization needs in scientific instruments and briefly reviews our progress leveraging the Chisel hardware description language and emerging open-source hardware ecosystems, including a few design examples.


\section{Introduction}

As CMOS scaling is coming to an end, specialization and heterogeneity are becoming crucial factors for sustaining performance growth, increasing energy efficiency, and improving resource utilization of computing hardware. Until recently, the shrinking of CMOS transistors has masked the performance overheads associated with general-purpose architectures.
Additionally, a perpetually evolving rich software ecosystem has accelerated software development on general-purpose architectures. Together, these trends have delayed the need for custom hardware solutions, which tend to be relatively expensive. However, with the end of the era of exponential performance growth and with the inefficiency associated with general-purpose architecture becoming increasingly made manifest~\cite{hameed2010understanding}, more specialized hardware, particularly AI accelerators, are emerging to fill the gap in computing environments~\cite{ovtcharov2015accelerating}.

Most AI success stories thus far have been associated with high-volume, mission-critical applications or high-profile projects in ``hyperscale'' companies such as Google, Microsoft, and Amazon. 
On the other hand, in the U.S. Department of Energy (DOE) scientific space, hardware customization needs are highly domain-specific (e.g., X-ray detectors) and thus relatively low volume (e.g., single unit to hundreds of units manufactured). Unfortunately, such low-volume use cases are not commercially lucrative for hardware vendors. Moreover, in many cases, such domain-specific hardware development tends to produce one-off implementations that are unusable beyond the initial intended scope.
What is needed is to identify common hardware building blocks (e.g., compressor, encryption blocks) that can broadly cover our scientific edge-compute needs.
We envision the co-design of hardware and software algorithms that can efficiently generate such modular and reusable components, and we develop a rich set of highly parameterized open-source hardware libraries. While our idea bears some resemblance to ``intellectual property'' (IP) cores provided by ASIC and FPGA vendors in terms of reusability, our goal is to develop hardware libraries of algorithmically complex blocks that, by virtue of their parameterization, are vendor and architecture agnostic. To this end, it is paramount to investigate (1) a hardware design environment that can capture practical high-level hardware design patterns (i.e., hardware algorithms) and can flexibly express them, (2) a lightweight, open (no commercial license) environment, including fast simulation and verification,  for hardware libraries (to accelerate the design loop and maximize distributed development), and (3) a lightweight resource estimation tool. While the end-to-end chip development cycle is important eventually, we focus primarily on the development cycle of front-end ASIC designs including digital circuit implementation, functional verification, and simulation. At the feasibility study phase, the crucial first step is to investigate the implementabibility of hardware algorithms for our hardware specialization needs (e.g., on-chip streaming processing logics). Additionally,  many ongoing research efforts are working to reduce the complexity of ASIC back-end design cycles using AI techniques such as floor planning~\cite{kahng2021ai}.

This manuscript is organized as follows. In Section~\ref{sec:backgrounds} we first discuss our motivation for edge hardware specialization. We then discuss the current status, challenges, and opportunities for hardware specialization; and we describe the challenges and opportunities in hardware specialization for the edge in the realm of scientific instruments.
In Section~\ref{sec:programming} we discuss hardware abstraction and programming languages.In Section~\ref{sec:workflow} we illustrate our proposed workflow for our hardware library approach with our X-ray pixel detector's data compressor block as an example. In Section 6 
 we discuss our research directions and future needs in hardware specialization and approaches for hardware co-design in the post-Moore era.

\section{Background}
~\label{sec:backgrounds}
\begin{figure}[tb]
\centering 
\includegraphics[width=.8\textwidth,origin=c,angle=0]{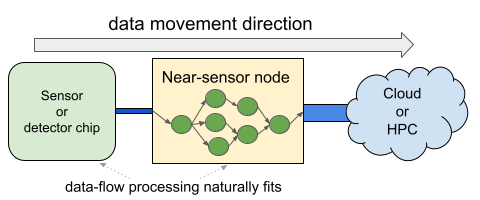}
\vspace*{-6mm}
\caption{Dataflow architecture opportunity at the edge of science.}
\vspace*{-5mm}
\label{fig:edgedataflow} 
\end{figure}

In a wide range of scientific applications, a data acquisition system
(possibly CPU or FPGA) collects values from sensor systems through
a low-level communication protocol (e.g., I2C, SPI, parallel and serial
buses) and transmits them to high-performance computing (HPC) systems
over traditional high-speed networks. The sensor systems
can range from simple individual sensors to more complex sensors paired
with some intelligence extended by microprocessors, for example, FPGAs and ASICs. 
In general, this dataflow is predominantly unidirectional, as illustrated in Figure~\ref{fig:edgedataflow}.

While various phases of sensor data analysis  can be lossy, 
it is crucial to transfer data from sensors to cloud or HPC systems without any reduction in data quality. 
This seemingly straightforward constraint is becoming a stiff challenge for many scientific instruments
because of the explosion in the volume of data produced by scientific
instruments as their spatial and temporal resolutions have continued to increase.
For example, the highest frame rate of X-ray pixel detectors will
soon approach a megahertz, necessitating an interconnect capable of
transferring a terabit of raw data off the chip. Such an interconnect
is cost-prohibitive. Other scientific disciplines, such as high-energy
physics, cosmology, and environmental remote sensing, are also facing the
same data bottleneck challenge. General-purpose processors at or near sensor nodes are no longer able to keep up with
this increase in data rate since they are burdened with unnecessary inefficiency~\cite{hameed2010understanding}
due to the nature of load-store architectures, highly complicated
memory subsystems, and lack of directional I/O connectivity. 
Note that novel accelerators, such as neuromorphic architectures, will not solve such problems for similar reasons. Adding larger on-chip memories that store captured data temporarily can be an intermediate solution; however, memory capacity is finite, which limits the time of experiments, and ultimately the solution is cost-ineffective. Thus, we need streaming (ideally stall-free) data-processing hardware designs that rely less so on temporary storage.

We believe a higher degree of hardware specialization (e.g., dataflow architecture) will be required to enable and support future scientific instrument needs.
For example, for X-ray pixel detectors,
the goal is to develop on-chip data compression algorithms
that compress data and can be placed directly adjacent to the internal sensors in the detector ASIC~\cite{Hammer_2021}. Since our hardware
development resources are currently limited, we focus on
co-designing a simple, yet effective, lossless compressor that
leverages application-specific characteristics and hardware specialization.
This reduces the complexity of implementation. Moreover, 
to minimize the hardware development efforts (particularly hardware
verification), we employ stateless dataflow architecture techniques
wherever possible. In addition to being ideal with respect to both
throughput and energy efficiency, such architectures are a natural
fit in the case of scientific instruments at the edge. We are also
conducting a feasibility study on embedding AI classification components
directly in the detector chip, which can be seen as a form of maximally
efficient data compression. This work has led to several interesting and critical 
research questions. These include the following: Which hardware architectures or platforms
(FPGA, ASIC, structured ASIC, coarse-grained reconfigurable architecture, etc.) are suitable for our future
needs? Which hardware programming languages or models improve our
productivity and support our innovations? Which architectural techniques
and design patterns are effective? How many components can be
expressed as stateless dataflow designs? Considerable research and development are required to answer the questions. 


Historically, hardware design ecosystems for ASICs have been dominated
by a few companies, and the licensing fees for their commercial
tools can often be a significant barrier to entry for many organizations.
Additionally,
the DOE scientific community is bifurcated: either projects have little
experience in hardware design, or they may have internal hardware designers
as part of their larger team who focus on single-purpose and project-/scope-optimized 
designs. As we are entering the post-Moore era, we believe
hardware specialization is the only practicable approach to dealing
with scaling problems, irrespective of the challenges awaiting us.
 We stress, however, that the number of hardware designers, developers, and architects is considerably low---orders of magnitude smaller than that of software developers. We believe that a lightweight and ergonomic development environment may help attract more developers
and students and enable cross-training of current software developers as hardware practitioners. 

The exploration of hardware programming ecosystems (e.g., programming models, tools) has been gaining significant attention in the field of computer architecture~\cite{truong2019golden,hennessynew} recently. This trend has also been the catalyst for an open-source hardware ecosystem boom, including instruction sets such as RISC-V~\cite{asanovic2014instruction}, hardware implementations~\cite{celio2017boomv2,fatollahi2014opensoc}, and  hardware tools~\cite{wolf2013yosys,Bachrach:ft,izraelevitz2017reusability,snyder2013verilator}.
Indeed, Google recently announced an open-source foundry PDK for SkyWater's 130 nm node~\cite{ansell2020missing}. Furthermore, with the advent of innovative companies such as efabless~\footnote{An open community for analog and mixed-signal IC and IP development and commercialization~\url{https://efabless.com/}}, even a small R\&D unit can now tape-out a chip with significantly lower cost, benefiting from a fully open-source and end-to-end ASIC design flow~\cite{edwards2021real}.
Additionally, we have been observing an interesting connection between software and hardware in terms of abstraction. Although no single abstraction can capture all features of a solution,  multiple abstraction layers, with each layer bridging adjacent layers, are needed for developing a complex system efficiently. 
While abstraction is, in general, well studied in the software world, such has  not been the case in the hardware world---but we observe that this now changing. 
Indeed, the hardware community is learning and incorporating ideas from software. 
Thus, we  believe that now is the best and most exciting time to participate in and contribute to the open-source hardware ecosystem.

Productivity, which may include debugability, maintainability, and reusability, is of the utmost concern for the DOE scientific community. Therefore,
investigation and development of efficient hardware abstractions
and productive hardware languages are of utmost importance to the community~\cite{truong2019golden}.
Many research groups in DOE have explored high-level synthesis (HLS), which
translates software codes into hardware description language (HDL) codes, such as OpenCL~\cite{czajkowski2012opencl},
in order to run high-performance computing kernels on FPGAs. This
approach does attract software developers and has shown promising results on particular kernels or parameters.
The downside of HLS is that optimization still requires hardware knowledge and offers no control over the generated HDL
codes. Small changes in the HLS source code will often lead to large changes in the generated HDL codes, making debugging almost impossible. It also incurs tighter dependencies on the underlying platforms and tool versions, which negatively affect maintainability and reusability. 
HDL, particularly Verilog, is still the main workhorse for I/O designs and other low-level hardware designs. A possible practical compromise may be to develop a domain-specific hardware language for scientific instruments at the edge, 
stitching HDL, HCL, and HSL together and providing a higher-level domain-specific abstraction.
While we primarily discuss Chisel in this paper, several other modern HDLs do exist and merit further investigation: MyHDL~\cite{decaluwe2004myhdl} and Migen~\cite{migen} for Python, and  C$\lambda$asH~\cite{essay59482} for Haskell. Even in the Scala language, there are other HDLs: SpinalHDL~\cite{spinalhdl}, which is essentially a ``fork'' of Chisel, and Spatial~\cite{koeplinger2018spatial}, whose chief improvement over Chisel is its incorporation of polyhedral compilation functionality.

\section{Hardware Programming Ecosystem}
\label{sec:programming}

Hardware description languages~\cite{golson2016language},
such as Verilog and VHDL, are used to describe digital circuits. Both
Verilog and VHDL were initially developed (in the early 1980s) for
digital circuit simulation. These languages eventually incorporated synthesis functionality and are currently ubiquitous in both industry and academia.
Although HDLs still play a primary role in digital circuit designs, as the complexity of hardware algorithms increases, the productivity of these traditional HDLs is becoming a matter of huge concern because of their lack of expressive language features, such as those found in modern general-purpose programming languages. 

One of the biggest problems with HDLs is that their mechanisms for generating recursive
(or tile) structures are weak. Thus, designers have to manually
``unroll'' modules with manual labeling and thereby reduce the codes' conciseness, maintainability,
and reusability. Examples of this kind of rolling can
be seen in any standard Verilog or VHDL implementation of a large
barrel shifter (i.e., a 2D array of multiplexers), systolic array~\cite{kung1978systolic}, network-on-chip architecture~\cite{kwon2017opensmart}, or variable-size
and fixed-size data converter~\cite{Ueno:2017hv}, the latter being
critical for data compression. To compensate for the weakness in recursive
generating capability, designers tend to use a general-purpose programming language, such as Python, C++, or TCL, to build generators that themselves generate HDL codes. 
Unsurprisingly, these kinds of toolchains are quite rigid (i.e., single-purpose), often brittle, and not scalable. 
Furthermore, since HDL design predates modern software
paradigms such as functional programming and test-driven development, as the complexity of software references or models increases, so do the
challenges associated with developing ``testbenches'' that precisely 
exercise designs implemented as HDL codes. Thus we
seek a hardware description language that enables a highly productive
development process such that we can use it to describe a circuit concisely,
express recursive structure, and describe testbenches more flexibly, among other features. 

\subsection*{Chisel hardware construction language}

Inspired by the RISC-V agile development approach~\cite{lee2016agile},
we chose an emerging hardware ``construction'' language named Chisel~\cite{Bachrach:ft,bachrach2017chisel,schoeberl2020digital}
for implementing our hardware libraries, performing functional simulations,
and designing extensions of our data compression block. Chisel,
which
stands for \textit{Constructing Hardware in a Scala Embedded Language}, is
designed to accelerate the digital circuit design process. As the name implies, Chisel is an embedded domain-specific language implemented
as a class library in Scala~\cite{odersky2008programming}. Chisel
offers a zero-cost abstraction of digital circuits, which means the overhead in performance
and resource usage induced by the abstraction is nearly zero
compared with that of a native HDL. Several studies have confirmed that Chisel significantly reduces the code size, improves code reusability, and incurs little performance
penalty compared with native Verilog implementations (i.e., for most
cases, Chisel-generated register-transfer level and equivalent
native Verilog implementation run at the same frequency~\cite{arcas2014empirical,mosanu2019flexi,kwon2017opensmart}).
We note that these  results are in stark contrast to 
HLS tools  that let developers specify design
behavior in higher-level languages and then infer lower-level
implementations. Such synthesis tools incur the cost of higher performance overheads,
particularly in resource usage.\footnote{The performance offered by a carefully optimized HLS implementation
is comparable to that of an HDL implementation; however, resource
usage is still in question.}

Leveraging the power of Scala's modern programming language features,
Chisel offers higher expressivity, which dramatically improves the productivity and flexibility for constructing synthesizable digital circuits. It also integrates
cycle-accurate simulators seamlessly, aiming for lightweight test-driven
development, which significantly reduces the barrier to entry for hardware development. 
In essence, Chisel has brought modern software paradigms to hardware
development, which should attract more software developers
to hardware development, thereby growing the community and ecosystem.
Indeed, since Chisel's original release in 2012~\cite{Bachrach:ft}, the
Chisel community has grown steadily, with the number of Chisel-based
open-source projects increasing every year. Chisel is being used for
many real-world tape-out designs~\cite{Asanovic:EECS-2016-17,celio2017boomv2,bailey201828nm,cass2019taking}
such as SiFive's RISC-V cores
and Google’s TPU~\cite{googletpu}. Chisel has also become popular
in academia for architecture research~\cite{di2018parallel,serre2018dsl,nowatzki2017stream,prabhakar2017plasticine}. 
Additionally, it has a rich ecosystem. 
For example, Chipyard~\cite{amid2020chipyard}, a framework for developing systems-on-chip (SoCs), encompasses an in-order core generator called Rocket Chip~\cite{rocketchip}, an out-of-order core generator named  BOOM~\cite{asanovic2015berkeley}, hardware accelerators such as Gemmini~\cite{genc2019gemmini}, and a systolic array generator~\cite{genc2019gemmini}.
Additionally, ChiselDSP~\cite{chiseldsp}, a library for generating more traditional signal-processing designs such as fast Fourier transforms, is also available.
Furthermore, ChiselVerify~\cite{dobis2021opensource} is a verification tool that employs industry-standard Universal Verification Methodology to verify Chisel designs formally.

To compare and contrast Chisel and Verilog, we
first consider a simple counter circuit written in Verilog and Chisel.
Listing \ref{verilog} is a Verilog implementation of a counter that
increments every cycle, counting from 0 to 9.
Listing \ref{chisel} is the same circuit written in Chisel. This example
highlights some fundamental feature differences between Chisel and Verilog. First,
the Chisel version has no clock, reset signals, or \texttt{always}
blocks; it automatically incorporates clock and reset signals when generating
Verilog codes. \texttt{RegInit}
creates a register that is initialized on reset with a specified value
(in this example, 0 as an \texttt{nbits}-wide 
unsigned integer); and thus, implicitly, a reset signal is generated.
Assigning to the register value \texttt{cntReg}
is translated to a nonblocking assignment via an \texttt{always}
block in Verilog. Since the default policy of the state element provided
by Chisel is a positive-edge register that supports synchronous reset,
no \texttt{always} blocks
are needed. Such a default policy in the state element enforces a
design guideline transparently and makes Chisel syntax more concise,
at the cost of flexibility. Chisel provides frequently used data types
such as \texttt{Int}, \texttt{UInt}, and \texttt{SInt},
instead of a range of bits as in Verilog, which also improves readability.
This Chisel code snippet also includes an example of parameterization;
the \texttt{Counter} class accepts the maximum value of the counter when instantiating and calculates the required bit length for the counter using \texttt{log2Ceil}.

\begin{lstlisting}[language=verilog,caption={A simple counter in Verilog. Note that \$clog2 is supported by SystemVerilog or Verilog-2005},captionpos=b,label={verilog},frame=single]
module Counter
  #(parameter MAXCNT=9)  (
  input        clock,
  input        reset,
  input        enable,
  output [$clog2(MAXCNT+1)-1:0] out
);
  reg [$clog2(MAXCNT)-1:0] cntReg;
  assign out = cntReg;
  always @(posedge clock) begin
    if (reset) begin
      cntReg <= 0;
    end else if (enable) begin
      if (cntReg == MAXCNT) begin
        cntReg <= 0;
      end else begin
        cntReg <= cntReg + 1;
      end
    end
  end
endmodule
\end{lstlisting}

\begin{lstlisting}[language=scala,caption={A simple counter in Chisel},captionpos=b,label={chisel},frame=single]
class Counter(val max:Int = 10) extends Module {
  val nbits = log2Ceil(max+1)
  val io = IO(new Bundle {
    val enable = Input(Bool())
    val out = Output(UInt(nbits.W))
  })
  val cntReg = RegInit(0.U(nbits.W))
  cntReg := Mux(io.enable,
   Mux(cntReg === max.U, 0.U, cntReg + 1.U), cntReg)
  io.out := cntReg
}
\end{lstlisting}

An essential aspect of Chisel is that it encourages test-driven
development and offers fully integrated testing harnesses. This allows users to write testbench codes in Scala, and running testbenches requires
no additional hardware. Listing~\ref{testbench} is a simple testbench
for the counter circuit (Listing~\ref{chisel}). Since the language
for testbenches is Scala, we can leverage its general-purpose
programming features. Chisel's peek-poke-expect-step test harness
is powerful and intuitive to use. For example, testbenches for our
data compression components directly read and analyze massive X-ray datasets
and feed selected regions to our compressor designs running in a circuit
simulator to verify expected functionality. Such ergonomic testing
enables effective software/hardware co-design.
Chisel simulates the behavior of generated circuits  using either an
internal Scala-based simulator or Verilator~\cite{snyder2013verilator},
which translates Verilog codes into cycle-accurate models (specified
as C++ codes) for faster simulation. In our experience, Chisel's integrated
test accelerates iteration of the design loop (coding, compiling,
evaluating), which greatly aids our design exploration, even when I/O
layout and parameters frequently change during the exploration. 

\begin{lstlisting}[language=Scala, caption={Testbench for counter}, captionpos=b, label={testbench},frame=single]
class CounterUnitTester(c: Counter) extends PeekPokeTester(c) {
  val max = c.max
  var ref = 0 // software reference count
  def test(e : Int) {
    poke(c.io.enable, e)
    for (i <- 0 until max+2) {
      // comparing hardware with software reference
      expect(c.io.out, ref) 
      step(1) // forward a single cycle
      if (e==1) if(ref < max) ref += 1 else ref = 0
    }
  }
  test(1)  // enable counting
  test(0)  // disable counting
}
\end{lstlisting}



\begin{figure}[h] 
\centering 
\includegraphics[width=.7\textwidth,origin=c,angle=0]{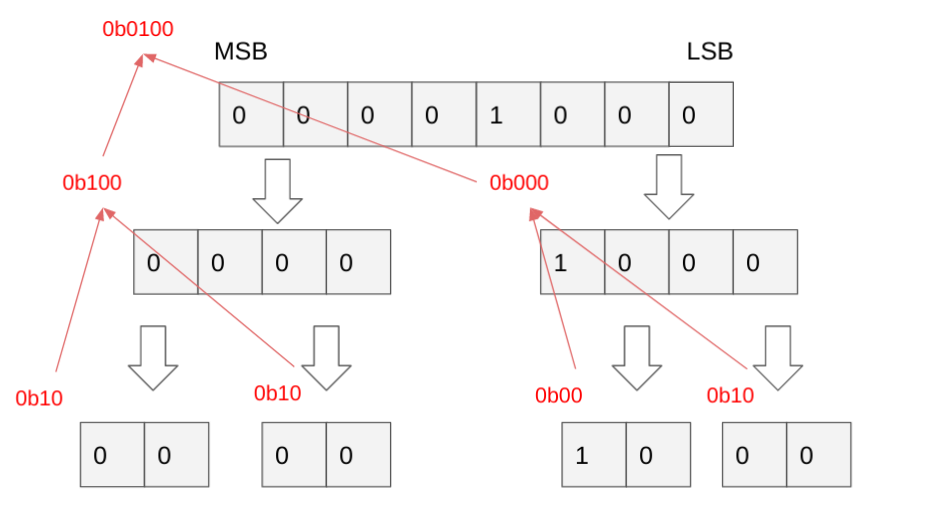}
\vspace*{-5mm}
\caption{Leading-zero counts divide-and-conquer algorithm.}.
\vspace*{-7mm}
\label{fig:dqclz} 
\end{figure}

A notably powerful feature of Chisel is its ability to construct
recursive hardware in a traditional software-like manner, such as divide-and-conquer. As 
mentioned, many hardware algorithms are hierarchical and recursive.
For example, leading zero counting can be implemented in
a divide-and-conquer manner (Figure ~\ref{fig:dqclz}). 
If we recursively split a binary input stream into two-bit words,
we can easily form the leading zero count with a look-up table (e.g.,
the leading zero count of 00 is 2). When two ``conquered'' blocks are merged, the leading zero counts of the merged block can be calculated from the leading zero counts of the conquered blocks, again with a simple bitwise XOR. 
One can implement the leading zero count in SystemVerilog using its ``generated'' primitive; however, conciseness and readability may be a concern.
On the other hand, Chisel enables specifying such a divide-and-conquer
algorithm directly and parameterizes the implementation.
Listing \ref{recursive} is a Chisel implementation example of leading
zero counting.

\begin{lstlisting}[language=Scala, caption={Recursive structure: counting leading zeros}, captionpos=b, label={recursive},frame=single]
class Clz(nb: Int = 64) extends Module {
  val lognb = log2Floor(nb)
  val io = IO(new Bundle {
    val in = Input(UInt(nb.W))
    val out = Output(UInt((lognb + 1).W))
  })
  if (nb == 1) io.out := !io.in
  else if (nb == 2) io.out := MuxLookup(io.in, 0.U, Array(0.U -> 2.U, 1.U -> 1.U))
  else {
    val largestPow2 = 1 << log2Ceil(nb) - 1
    val c0 = Module(new Clz(largestPow2))
    val c1 = Module(new Clz(nb - largestPow2))
    c0.io.in := io.in(nb - 1, nb - largestPow2)
    c1.io.in := io.in(nb - largestPow2 - 1, 0)

    io.out := Mux(c0.io.out(c0.lognb), largestPow2.U +& c1.io.out, c0.io.out)
  }
}
\end{lstlisting}





\section{Co-design Workflow for Hardware Libraries}
\label{sec:workflow}
In this section we first describe our current design workflow for
developing hardware libraries. Next we explain how we are applying the
workflow to our scientific instrument edge-computing using a data reduction 
stage for an X-ray pixel array detector ASIC chip. One of our requirements is that 
the workflow be easy to deploy to a typical development
environment, such as a Linux server or laptop, without requiring special software licensing. In fact, ideally, the workflow consists of open-source
tools. 
Other than for philosophical reasons, the additional motivation for 
the easy-to-install and open-source workflow is to attract more hardware 
developers and students and  to enable the training of software developers in hardware design.  
This is due to the fact that conventionally trained hardware developers remain scarce and there is no guarantee that this situation will improve in the near to medium term.

As we described, Chisel is fully open-source software that offers
sufficient features for efficiently describing algorithmically complex
hardware designs and provides a flexible testbench framework that
is seamlessly integrated with a typical workflow. At this point, however,
Chisel lacks lightweight resource estimation functionality (e.g., counting
the number of logic gates, wires). Since both ASIC and FPGA are spatially
constrained, the resource usage of a hardware design must be bounded
by available resources. Unlike software platforms, hardware in general
offers no dynamic memory allocation, time sharing or context switching,
or virtualization of hardware logic. Thus, estimating resource usage
for the target range of hardware parameters is crucial for hardware
libraries and is one the most important steps of co-design.

Accurate estimation of hardware resource usage can be complicated, however,
requiring domain expertise; moreover, depending on the platforms or technology,
it may require prohibitively expensive commercial tools. For our purposes of
estimating realistic resource usage for ASICs,~\footnote{Currently, only a select few technology nodes can be targeted with open-source EDA tools.} we need only count the number of basic components such as wires, flip-flops,
logic gates, and multiplexers efficiently. Several open-source
digital circuit simulators or synthesis tools can give us such a resource
estimate. We found that Yosys~\cite{wolf2013yosys}, an open-source
synthesis tool for ASICs and FPGAs, provides a statistical report on
resource usage for both ASIC and FPGA. Since Yosys is a full-featured
RTL synthesis tool, 
it can also perform synthesis-level optimizations,
such as removing redundant multiplexers, which can give us a more
realistic resource usage estimation (since, in fact, all synthesis tools
perform optimizations). 



\subsection*{Compressor/Reduction Logics}
To illustrate our co-design workflow, we present a conceptual overview of our data compressor hardware designs for X-ray detectors, which will be placed in our X-ray detector chip that we are currently developing. To simplify and focus on the main points, we exclude the details of our hardware design that are specific to our X-ray array detector and its I/O characteristics.

\begin{figure}[tb]
  \begin{minipage}[b]{0.47\textwidth}
    \includegraphics[width=\textwidth]{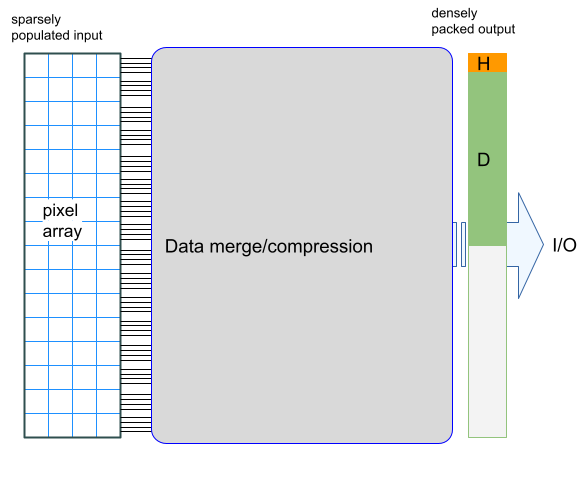}
    \vspace*{-5mm}
    \caption{Stream compressor concept (converts a sparsely populated input into  densely packed, compressed data)}
\vspace*{-5mm}
    \label{fig:pad-comp-concept} 
  \end{minipage}
  \hspace{0.6cm}
  \begin{minipage}[b]{0.47\textwidth}
    \includegraphics[width=\textwidth]{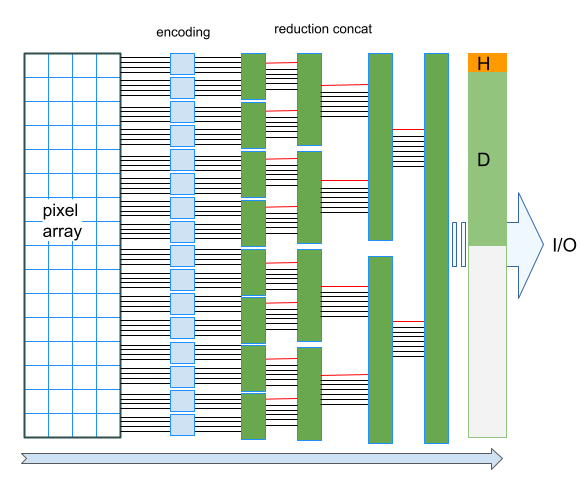}
    \vspace*{-5mm}
    \caption{Our design approach (concatenates variable-sized encoded outputs with hierarchical reduction stages).}
\vspace*{-5mm}
    \label{fig:pad-comp-approach} 
  \end{minipage}
\end{figure}

The pixel array generates data every cycle, where the total number
of pixels generated per cycle depends on the hardware design parameters
(e.g., 1,024 pixels or 8 columns $\times$ 128 rows  of pixels). Data
from an X-ray pixel array is generally sparse and contains few nonzero
pixels. Our previous analysis of real datasets showed that a large percentage
(e.g., 80\% or greater) of X-ray pixel data have zeros (or even
lower values) that occupy only the lower few bits. Hence, the purpose of our compressor block is to compress sparsely populated input data
into a densely packed stream that includes a header and compressed data
without stalling, in order to reduce the total amount of data that
needs to be sent to the I/O block (Figure~\ref{fig:pad-comp-concept}).
For our detector chip, we must minimize the size of the internal buffer
to temporarily store pixel data because of two factors: (1) the size
of temporary memory limits the duration of experiments, and (2) memory
is scarce on the detector ASIC chip. Additionally, since the cost
of validation is large, dataflow designs with minimum state
elements are preferable. For this reason, any compression approach
based on an entropy algorithm such as Huffman coding~\cite{hashemian1994design}
may be unsuitable for our on-chip compressor.

Our data compression dataflow architecture consists of an encoding
stage and reduction stages (Figure~\ref{fig:pad-comp-approach}).
Since zero (or lower) values dominate the majority of the input data,
we employ a bit-shuffling scheme in the encoding stage, which resembles
a matrix transpose operation and increases the co-occurrence of zero
pixels, to filter out unused higher bits. The bit-shuffling operation
can be expressed simply as a set of wires between the input bits and
the output bits in the correct order and requires no logic circuit
in ASIC; hence it is inexpensive to implement and verify~\cite{Hammer_2021}.
The output from the encoding stage is variable in size, and multiple
encoding blocks generate variable-sized data simultaneously, so no I/O
can handle such inputs directly. We employ a reduction stage to concatenate variable-sized data into a single continuous block in a hierarchical manner.

\begin{figure}[tb]
  \begin{minipage}[b]{0.5\textwidth}
    \includegraphics[width=0.9\textwidth]{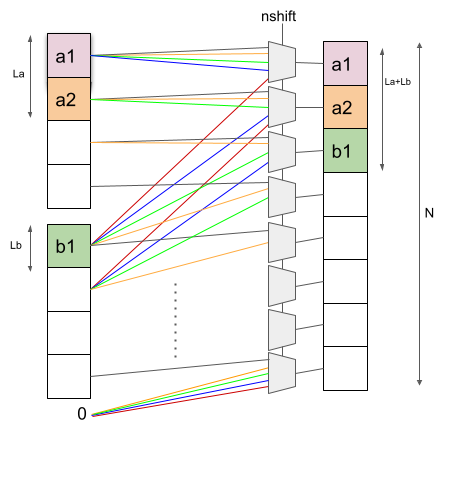}
    \vspace*{-5mm}
    \caption{Concat baseline implementation.}
\vspace*{-5mm}
    \label{fig:concatnaive} 
  \end{minipage}
  \begin{minipage}[b]{0.5\textwidth}
    \includegraphics[width=0.9\textwidth]{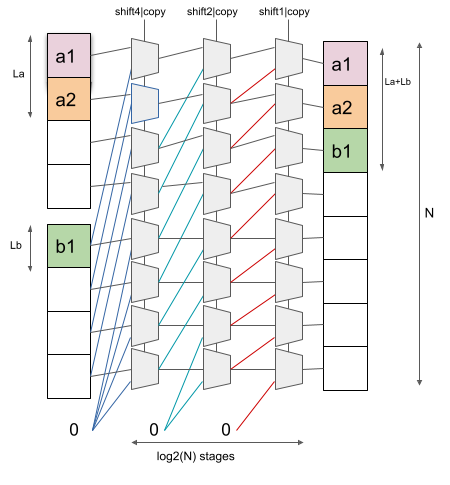}
    \vspace*{-5mm}
    \caption{Concat optimized implementation.}
\vspace*{-5mm}
    \label{fig:concatnlogn} 
  \end{minipage}
\end{figure}

Figure~\ref{fig:concatnaive} illustrates a baseline implementation, which uses
multiple-input multiplexers to select input pixels. The input consists
of two fixed data arrays, where the size is $N$ and the length of
the content is $L_{a}$ and $L_{b}$, respectively. The size of
the output array is $2N$, and the length of the concatenated content
is $L$. This implementation approach requires $\left(N^{2}+N\right)/2$
multiplexers. The baseline approach is straightforward to implement and can be implemented
only with combinational logic for ASICs,\footnote{FPGA deployments may require pipelining for the sake of achieving low latency.}
but the resource consumption could be prohibitive for larger inputs.
Figure~\ref{fig:concatnlogn} shows an optimized implementation that
shifts the second input using a series of shifting stages, where each
stage shifts only a specific power of 2. The required number
of shifting stages is $\log\left(N\right)$, where $N$ is the size
of each input array. If the size of each input array is 4, for example,
three shifting stages are needed (shifting 4 pixels, 2 pixels, and
1 pixel). The expected number of multiplexers in this implementation
is $N\log\left(N\right)$. Both hardware algorithms can be easily
implemented in Chisel and fully parameterized, including the testbenches,
thanks to the power of modern language features such as functional
programming.

The concision of the Chisel implementation improves the maintainability
and helps other developers understand the fundamentals of the design,
 a crucial concern for hardware libraries. Table~\ref{tbl:loc}
includes the number of lines of code (LOC) for each algorithm in Chisel
and the LOC of a generated Verilog code with a specific
target size. Since the Chisel implementation is fully parameterized,
including testbenches, the number of the LOC\footnote{significant lines of code, i.e., excluding comments} is the same for any input size.

\begingroup
\setlength{\tabcolsep}{8pt}
\renewcommand{\arraystretch}{1.5}
\centering
\begin{table}[h!]
 \caption{Number of lines: Chisel and generated Verilog code}
 \begin{tabular}{|c | c | c |c  | c|} 
 \hline
   &  Chisel & Verilog (N=8) & Verilog (N=32) &  Verilog (N=128) \\ [0.5ex] 
 \hline\hline
 Baseline       & 54    & 432  & 3864  & 52152 \\ 
 \hline
 Optimized & 97    & 432  & 2250  & 11628 \\
 \hline
\end{tabular}
\label{tbl:loc}
\end{table}
\endgroup

\begin{figure}[tb]
  \begin{minipage}[b]{0.47\textwidth}
    \includegraphics[width=\textwidth]{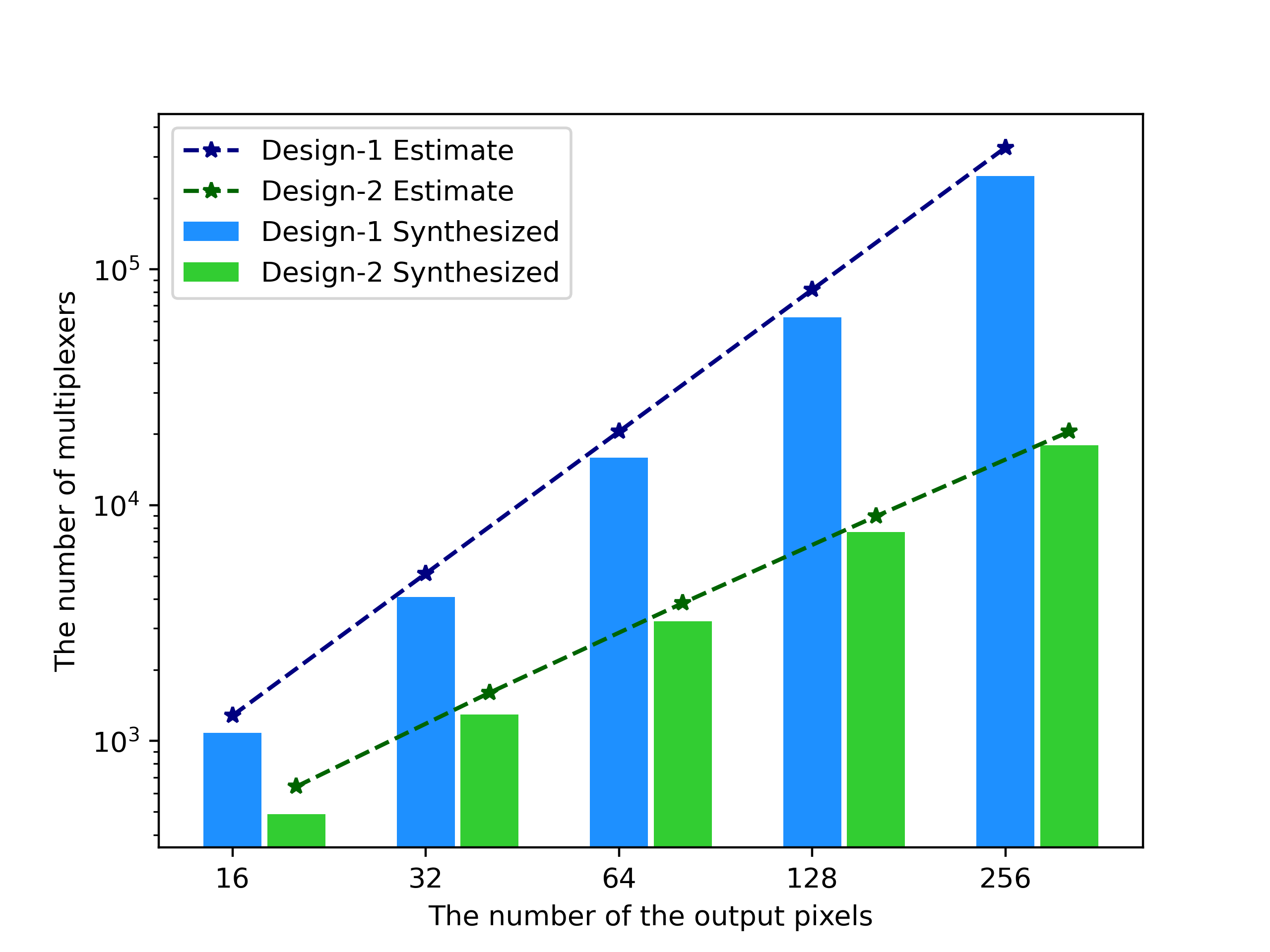}
    \vspace*{-5mm}
    \caption{ASIC multiplexer usage estimation with Yosys.}
\vspace*{-5mm}
    \label{fig:mux} 
  \end{minipage}
  \hspace{0.6cm}
  \begin{minipage}[b]{0.47\textwidth}
    \includegraphics[width=\textwidth]{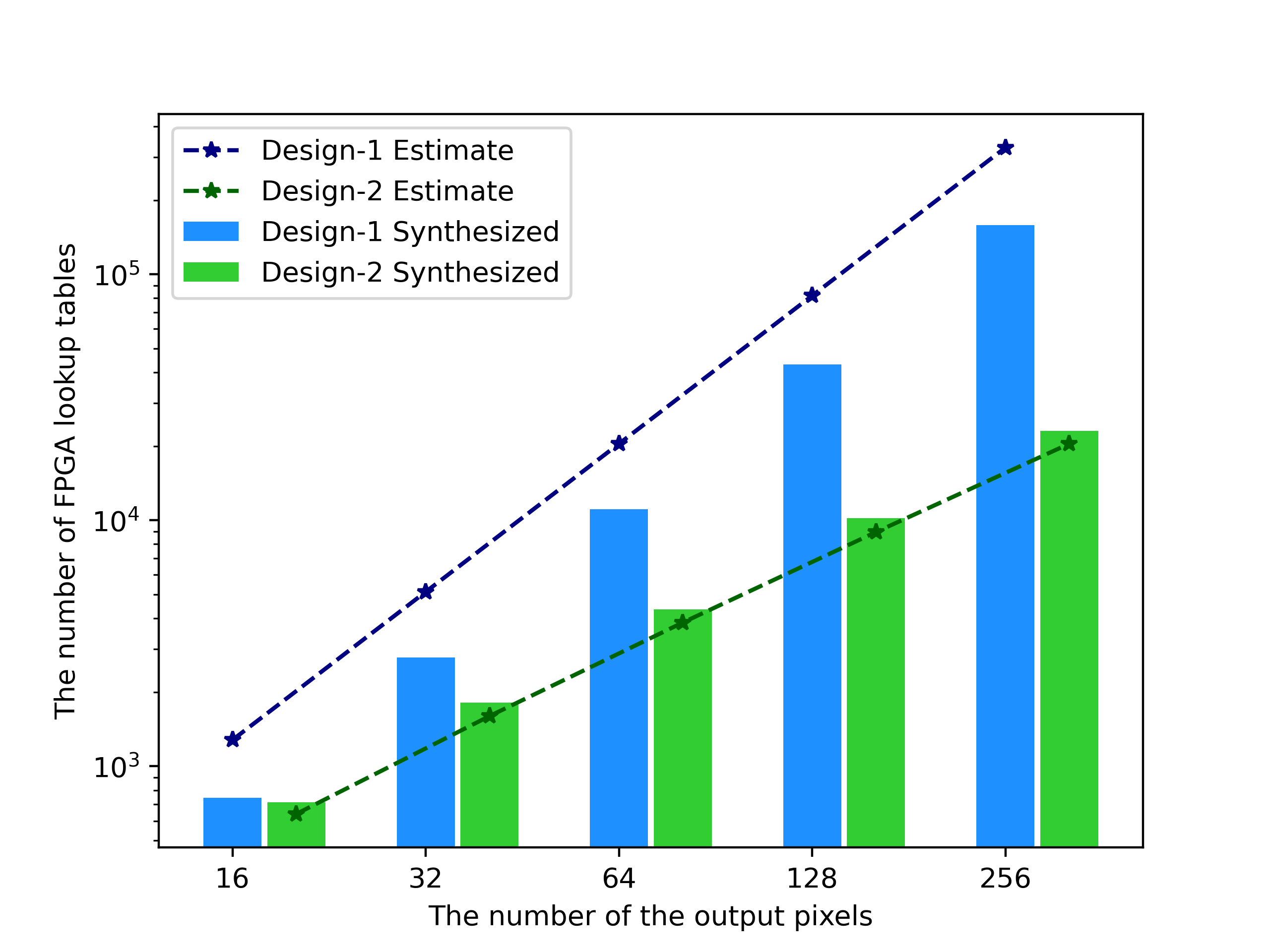}
    \vspace*{-5mm}
    \caption{FPGA lookup table usage estimation with Yosys.}
\vspace*{-5mm}
    \label{fig:lut} 
  \end{minipage}
\end{figure}

One important co-design criterion is resource usage. With a resource
estimate, we can discuss the feasibility of an RTL code generated from
a hardware library generator or the number of copies we can fit on a target hardware platform. For
the above example designs, the resource usage can be easily computed
analytically, although in general this may not be possible or may require extra effort. As a demonstration, Figure~\ref{fig:mux} and Figure~\ref{fig:lut} include a comparison of analytically predicted resources and a statistical report from Yosys. They compare the number of ASIC multiplexers and the number of FPGA lookup tables (LUTs) for the two designs. The analytical method reasonably captures the trendline of the resource usage for both ASIC and FPGA. Depending on the algorithms or underlying architecture, however, the room for optimizations varies. For the baseline algorithm, the gap between the analytical prediction
and the Yosys report is large.
The reason is that  synthesis tools, including Yosys, optimize multiplexer usage by removing unnecessary multiplexers and the redundancy of multiplexers in the baseline implementation artificially inflates multiplexer usage. 
FPGA-configurable logic blocks are also primitive building
blocks and generally contain flip-flops, multiplexers, and LUTs with
multibit inputs, which are much more complex than in the case of simple
multiplexers (e.g., 1-bit, 2-input). In terms of estimation time,
Yosys takes only up to a couple of seconds to report the resource
usage for fairly large implementations. Since Yosys is a command-line
tool, it is straightforward to integrate into our workflow.


\subsection{Lightweight AI Capability for Future Detector Systems}

While a full-fledged AI capability is impractical to implement on-chip, specialized, lightweight AI capabilities are worth consideration as a possible replacement for conventional methods such as data reduction by feature detection. Implementation of AI algorithms as hardware algorithms is a novel area of research, enabled by this growth in access to EDA tools.
Work in this area can be partitioned into two subareas, roughly along the dimension of generality. The first, and more general, involves the design and testing of domain-specific accelerators (e.g., GEMM accelerators such as Gemmini~\cite{genc2019gemmini} and NVDLA~\cite{nvdla}) as custom instruction set architectures (ISAs). These ISAs are conceptually not so different from general-purpose compute architectures (in that they are programmable) except insofar as they prioritize a limited number of operations, particularly those pertinent to AI workloads (such as matrix multiplication). Their value proposition to users is that very prioritization, leading to outsized efficiency and increased performance (on AI workloads). This work in the research community is mirrored in industry, where recently there has been a renewed ``Cambrian explosion'' in architecture companies (such as SiFive, Cerebras, Habana, and Samba Nova) that aim to support the same such AI workflows. The second subarea involves the design and testing of particular use-case-specific circuits, wherein a circuit is in one-to-one correspondence with, for example, a particular neural network. These implementations differ from those of the former in that they are not reconfigurable and represent, essentially, in silico instantiations of neural networks. The purported advantage of this approach is the hyper-refinement possible due to the specificity of the datapath and compute workload. Research in this area traditionally has employed HLS tools as ``compiler'' intermediaries between high-level implementations and the actual hardware design~\cite{Zhang2021fracbnn}, which suffer from unpredictable resource usage. Recently exploratory work has  been done using Chisel as the high-level implementation language \textit{and} the low-level description language~\cite{eldridge2015}. While this still is in its initial stages, we have begun to explore and develop such a solution for the purposes of Bragg peak detection~\cite{liu2020braggnn}.

\section{Conclusion}

 Hardware specialization is poised to play a critical role in the post-Moore era,
in improving both the performance and energy efficiency of computing
in general. In scientific edge computing in particular, since the volume of data generated by sensors is expected to increase exponentially,
placing computing power as close as possible to the data sources---where
general-purpose processors or accelerators fail to meet our requirements--- is becoming a crucial enabling factor for future scientific experiments.
Custom hardware development needs tend to
be confined to low-volume applications, with designs having low reusability.
To address these issues, we employed the Chisel hardware construction language
and other open-source design tools to create a co-design workflow
for exploring hardware algorithms and developing reusable hardware
libraries that can capture the design patterns of common hardware
components. In particular, dataflow components can accelerate our specialized hardware development for future scientific edge-computing applications. We described our open-source-tools-based co-design workflow and delineated the development of our data compression block for our X-ray detector ASIC chip. Productive hardware ecosystems could shape future hardware innovation, and we believe that we are entering the golden age of hardware programming. We
continue to investigate cutting-edge hardware ecosystems to improve
the productivity of hardware algorithms and libraries development on data compression, encryption, and AI classification, for scientific instrument edge-computing.

\section*{Acknowledgments}

We thank Ian Foster and Kyle Chard for supporting this exciting collaboration between Argonne National Laboratory and the University of Chicago Department of Computer Science. We thank Pete Beckman and Alec Sandy for encouraging this multidisciplinary collaboration between the X-ray Science Division (XSD) and the Mathematics and Computer Science (MCS) Division at Argonne National Laboratory. We also thank two anonymous referees for their  useful comments. We thank Gail Pieper for editing this manuscript. The material is based upon work supported by Laboratory Directed Research and Development (LDRD 2021-0072) funding from Argonne National Laboratory, provided by the Director, Office of Science, of the U.S. Department of Energy under contract DE-AC02-06CH11357.

\bibliographystyle{ieeetr}
\bibliography{refs}

\begin{thebibliography}{10}

\bibitem{hameed2010understanding}
R.~Hameed, W.~Qadeer, M.~Wachs, O.~Azizi, A.~Solomatnikov, B.~C. Lee,
  S.~Richardson, C.~Kozyrakis, and M.~Horowitz, ``Understanding sources of
  inefficiency in general-purpose chips,'' in {\em Proceedings of the 37th
  annual international symposium on Computer architecture}, pp.~37--47, 2010.

\bibitem{ovtcharov2015accelerating}
K.~Ovtcharov, O.~Ruwase, J.-Y. Kim, J.~Fowers, K.~Strauss, and E.~S. Chung,
  ``Accelerating deep convolutional neural networks using specialized
  hardware,'' {\em Microsoft Research Whitepaper}, vol.~2, no.~11, pp.~1--4,
  2015.

\bibitem{kahng2021ai}
A.~B. Kahng, ``Ai system outperforms humans in designing floorplans for
  microchips,'' 2021.

\bibitem{Hammer_2021}
M.~Hammer, K.~Yoshii, and A.~Miceli, ``Strategies for on-chip digital data
  compression for {X-ray} pixel detectors,'' {\em Journal of Instrumentation},
  vol.~16, pp.~P01025--P01025, jan 2021.

\bibitem{truong2019golden}
L.~Truong and P.~Hanrahan, ``A golden age of hardware description languages:
  applying programming language techniques to improve design productivity,'' in
  {\em 3rd Summit on Advances in Programming Languages (SNAPL 2019)}, Schloss
  Dagstuhl-Leibniz-Zentrum fuer Informatik, 2019.

\bibitem{hennessynew}
J.~Hennessy and D.~Patterson, ``A new golden age for computer architecture:
  Domain-specific hardware/software co-design, enhanced,'' {\em ACM/IEEE 45th
  Annual International Symposium on Computer Architecture (ISCA)}, 2018.

\bibitem{asanovic2014instruction}
K.~Asanovi{\'c} and D.~A. Patterson, ``Instruction sets should be free: The
  case for {RISC-V},'' {\em EECS Department, University of California,
  Berkeley, Tech. Rep. UCB/EECS-2014-146}, 2014.

\bibitem{celio2017boomv2}
C.~Celio, P.-F. Chiu, B.~Nikolic, D.~A. Patterson, and K.~Asanovic, ``{BOOMv2:
  an open-source out-of-order RISC-V core},'' in {\em First Workshop on
  Computer Architecture Research with RISC-V (CARRV)}, 2017.

\bibitem{fatollahi2014opensoc}
F.~Fatollahi-Fard, D.~Donofrio, G.~Michelogiannakis, and J.~Shalf, ``{OpenSoC}
  fabric: On-chip network generator: Using {C}hisel to generate a
  parameterizable on-chip interconnect fabric,'' in {\em Proceedings of the
  2014 International Workshop on Network on Chip Architectures}, pp.~45--50,
  2014.

\bibitem{wolf2013yosys}
C.~Wolf, J.~Glaser, and J.~Kepler, ``Yosys -- a free {Verilog} synthesis
  suite,'' in {\em Proceedings of the 21st Austrian Workshop on
  Microelectronics (Austrochip)}, 2013.

\bibitem{Bachrach:ft}
J.~Bachrach, H.~Vo, B.~Richards, and Y.~L. D. a. d.~. Design, ``{Chisel:
  constructing hardware in a {Scala} embedded language},'' {\em DAC Design
  Automation Conference}, pp.~1212--1221, 2012.

\bibitem{izraelevitz2017reusability}
A.~Izraelevitz, J.~Koenig, P.~Li, R.~Lin, A.~Wang, A.~Magyar, D.~Kim,
  C.~Schmidt, C.~Markley, J.~Lawson, {\em et~al.}, ``Reusability is {FIRRTL}
  ground: Hardware construction languages, compiler frameworks, and
  transformations,'' in {\em 2017 IEEE/ACM International Conference on
  Computer-Aided Design (ICCAD)}, pp.~209--216, IEEE, 2017.

\bibitem{snyder2013verilator}
W.~Snyder, ``Verilator: Open simulation-growing up,'' {\em DVClub Bristol},
  2013.

\bibitem{ansell2020missing}
T.~Ansell and M.~Saligane, ``The missing pieces of open design enablement: A
  recent history of {Google} efforts (invited paper),'' in {\em 2020 IEEE/ACM
  International Conference On Computer Aided Design (ICCAD)}, pp.~1--8, IEEE,
  2020.

\bibitem{edwards2021real}
R.~T. Edwards, M.~Shalan, and M.~Kassem, ``Real silicon using open-source
  {EDA},'' {\em IEEE Design \& Test}, 2021.

\bibitem{czajkowski2012opencl}
T.~S. Czajkowski, U.~Aydonat, D.~Denisenko, J.~Freeman, M.~Kinsner, D.~Neto,
  J.~Wong, P.~Yiannacouras, and D.~P. Singh, ``From {OpenCL} to
  high-performance hardware on {FPGAs},'' in {\em 22nd international conference
  on field programmable logic and applications (FPL)}, pp.~531--534, IEEE,
  2012.

\bibitem{decaluwe2004myhdl}
J.~Decaluwe, ``{MyHDL: a Python}-based hardware description language.,'' {\em
  Linux Journal}, no.~127, pp.~84--87, 2004.

\bibitem{migen}
``Migen, a {Python} toolbox for building complex digital hardware.''
  \url{https://github.com/m-labs/migen}.

\bibitem{essay59482}
C.~{Baaij}, ``C$\lambda$ash : from haskell to hardware,'' December 2009.

\bibitem{spinalhdl}
P.~Charles, ``Spinalhdl.'' \url{https://github.com/SpinalHDL/SpinalHDL}, 2016.

\bibitem{koeplinger2018spatial}
D.~Koeplinger, M.~Feldman, R.~Prabhakar, Y.~Zhang, S.~Hadjis, R.~Fiszel,
  T.~Zhao, L.~Nardi, A.~Pedram, C.~Kozyrakis, {\em et~al.}, ``Spatial: A
  language and compiler for application accelerators,'' in {\em Proceedings of
  the 39th ACM SIGPLAN Conference on Programming Language Design and
  Implementation}, pp.~296--311, 2018.

\bibitem{golson2016language}
``{}language wars in the 21st century: {Verilog versus VHDL}--revisited,''

\bibitem{kung1978systolic}
H.~T. Kung and C.~E. Leiserson, ``Systolic arrays for (vlsi).,'' tech. rep.,
  CARNEGIE-MELLON UNIV PITTSBURGH PA DEPT OF COMPUTER SCIENCE, 1978.

\bibitem{kwon2017opensmart}
H.~Kwon and T.~Krishna, ``{OpenSMART: Single-cycle multi-hop NoC generator in
  BSV and Chisel},'' in {\em 2017 IEEE International Symposium on Performance
  Analysis of Systems and Software (ISPASS)}, pp.~195--204, IEEE, 2017.

\bibitem{Ueno:2017hv}
T.~Ueno, K.~Sano, and S.~Yamamoto, ``Bandwidth compression of floating-point
  numerical data streams for {FPGA}-based high-performance computing,'' {\em
  ACM Transactions on Reconfigurable Technology and Systems}, vol.~10,
  pp.~1--22, July 2017.

\bibitem{lee2016agile}
Y.~Lee, A.~Waterman, H.~Cook, B.~Zimmer, B.~Keller, A.~Puggelli, J.~Kwak,
  R.~Jevtic, S.~Bailey, M.~Blagojevic, {\em et~al.}, ``An agile approach to
  building {RISC-V} microprocessors,'' {\em IEEE Micro}, vol.~36, no.~2,
  pp.~8--20, 2016.

\bibitem{bachrach2017chisel}
J.~J. Bachrach and K.~Asanovi{\'c}, ``Chisel 3.0 tutorial,'' {\em EECS
  Department, UC Berkeley, Tech. Rep.}, 2017.

\bibitem{schoeberl2020digital}
M.~Schoeberl, ``Digital design in {Chisel},'' 2020.

\bibitem{odersky2008programming}
M.~Odersky, L.~Spoon, and B.~Venners, {\em Programming in {Scala}}.
\newblock Artima Inc, 2008.

\bibitem{arcas2014empirical}
O.~Arcas-Abella, G.~Ndu, N.~Sonmez, M.~Ghasempour, A.~Armejach, J.~Navaridas,
  W.~Song, J.~Mawer, A.~Cristal, and M.~Luj{\'a}n, ``An empirical evaluation of
  high-level synthesis languages and tools for database acceleration,'' in {\em
  2014 24th International Conference on Field Programmable Logic and
  Applications (FPL)}, pp.~1--8, IEEE, 2014.

\bibitem{mosanu2019flexi}
S.~Mosanu, X.~Guo, M.~El-Hadedy, L.~Anghel, and M.~Stan, ``{Flexi-AES}: A
  highly-parameterizable cipher for a wide range of design constraints,'' in
  {\em 2019 IEEE 27th Annual International Symposium on Field-Programmable
  Custom Computing Machines (FCCM)}, pp.~338--338, IEEE, 2019.

\bibitem{Asanovic:EECS-2016-17}
K.~Asanović, R.~Avizienis, J.~Bachrach, S.~Beamer, D.~Biancolin, C.~Celio,
  H.~Cook, D.~Dabbelt, J.~Hauser, A.~Izraelevitz, S.~Karandikar, B.~Keller,
  D.~Kim, J.~Koenig, Y.~Lee, E.~Love, M.~Maas, A.~Magyar, H.~Mao, M.~Moreto,
  A.~Ou, D.~A. Patterson, B.~Richards, C.~Schmidt, S.~Twigg, H.~Vo, and
  A.~Waterman, ``The {Rocket} chip generator,'' Tech. Rep. UCB/EECS-2016-17,
  EECS Department, University of California, Berkeley, Apr 2016.

\bibitem{bailey201828nm}
S.~Bailey, J.~Wright, N.~Mehta, R.~Hochman, R.~Jarnot, V.~Milovanovi{\'c},
  D.~Werthimer, and B.~Nikoli{\'c}, ``A 28nm {FDSOI 8192-point digital ASIC
  spectrometer from a Chisel generator},'' in {\em 2018 IEEE Custom Integrated
  Circuits Conference (CICC)}, pp.~1--4, IEEE, 2018.

\bibitem{cass2019taking}
S.~Cass, ``Taking {AI to the edge: Google's TPU} now comes in a maker-friendly
  package,'' {\em IEEE Spectrum}, vol.~56, no.~5, pp.~16--17, 2019.

\bibitem{googletpu}
D.~Lockhart, S.~Twigg, R.~Narayanaswami, J.~Coriell, U.~Dasari, R.~Ho,
  D.~Hogberg, G.~Huang, A.~Kane, C.~Kaur, T.~Liu, A.~Maggiore, K.~Townsend, and
  E.~Tuncer, ``Experiences building edge {TPU with Chisel}.'' 2018 Chisel
  Community Conference, 2018.

\bibitem{di2018parallel}
L.~Di~Tucci, D.~Conficconi, A.~Comodi, S.~Hofmeyr, D.~Donofrio, and M.~D.
  Santambrogio, ``A parallel, energy efficient hardware architecture for the
  {merAligner on FPGA using Chisel HCL},'' in {\em 2018 IEEE International
  Parallel and Distributed Processing Symposium Workshops (IPDPSW)},
  pp.~214--217, IEEE, 2018.

\bibitem{serre2018dsl}
F.~Serre and M.~P{\"u}schel, ``A {DSL-based FFT hardware generator in Scala},''
  in {\em 2018 28th International Conference on Field Programmable Logic and
  Applications (FPL)}, pp.~315--3157, IEEE, 2018.

\bibitem{nowatzki2017stream}
T.~Nowatzki, V.~Gangadhar, N.~Ardalani, and K.~Sankaralingam, ``Stream-dataflow
  acceleration,'' in {\em 2017 ACM/IEEE 44th Annual International Symposium on
  Computer Architecture (ISCA)}, pp.~416--429, IEEE, 2017.

\bibitem{prabhakar2017plasticine}
R.~Prabhakar, Y.~Zhang, D.~Koeplinger, M.~Feldman, T.~Zhao, S.~Hadjis,
  A.~Pedram, C.~Kozyrakis, and K.~Olukotun, ``Plasticine: A reconfigurable
  architecture for parallel patterns,'' in {\em 2017 ACM/IEEE 44th Annual
  International Symposium on Computer Architecture (ISCA)}, pp.~389--402, IEEE,
  2017.

\bibitem{amid2020chipyard}
A.~Amid, D.~Biancolin, A.~Gonzalez, D.~Grubb, S.~Karandikar, H.~Liew,
  A.~Magyar, H.~Mao, A.~Ou, N.~Pemberton, {\em et~al.}, ``Chipyard: Integrated
  design, simulation, and implementation framework for custom {SoCs},'' {\em
  IEEE Micro}, vol.~40, no.~4, pp.~10--21, 2020.

\bibitem{rocketchip}
K.~Asanović, R.~Avizienis, J.~Bachrach, S.~Beamer, D.~Biancolin, C.~Celio,
  H.~Cook, D.~Dabbelt, J.~Hauser, A.~Izraelevitz, S.~Karandikar, B.~Keller,
  D.~Kim, J.~Koenig, Y.~Lee, E.~Love, M.~Maas, A.~Magyar, H.~Mao, M.~Moreto,
  A.~Ou, D.~A. Patterson, B.~Richards, C.~Schmidt, S.~Twigg, H.~Vo, and
  A.~Waterman, ``The rocket chip generator,'' Tech. Rep. UCB/EECS-2016-17, EECS
  Department, University of California, Berkeley, Apr 2016.

\bibitem{asanovic2015berkeley}
K.~Asanovic, D.~A. Patterson, and C.~Celio, ``The {Berkeley Out-of-Order
  Machine (BOOM)}: An industry-competitive, synthesizable, parameterized risc-v
  processor,'' tech. rep., University of California at Berkeley Berkeley United
  States, 2015.

\bibitem{genc2019gemmini}
H.~Genc, A.~Haj-Ali, V.~Iyer, A.~Amid, H.~Mao, J.~Wright, C.~Schmidt, J.~Zhao,
  A.~Ou, M.~Banister, Y.~S. Shao, B.~Nikolic, I.~Stoica, and K.~Asanovic,
  ``Gemmini: An agile systolic array generator enabling systematic evaluations
  of deep-learning architectures,'' 2019.

\bibitem{chiseldsp}
A.~Wang, {\em Agile Design of Generator-Based Signal Processing Hardware}.
\newblock PhD thesis, EECS Department, University of California, Berkeley, May
  2019.

\bibitem{dobis2021opensource}
A.~Dobis, T.~Petersen, K.~J.~H. Rasmussen, E.~Tolotto, H.~J. Damsgaard, S.~T.
  Andersen, R.~Lin, and M.~Schoeberl, ``Open-source verification with {Chisel
  and Scala},'' 2021.

\bibitem{hashemian1994design}
R.~Hashemian, ``Design and hardware construction of a high speed and memory
  efficient {Huffman} decoding,'' in {\em IEEE International Conference on
  Consumer Electronics}, pp.~74--75, IEEE, 1994.

\bibitem{nvdla}
F.~Farshchi, Q.~Huang, and H.~Yun, ``Integrating {NVIDIA Deep Learning
  Accelerator (NVDLA) with RISC-V SoC on FireSim},'' in {\em 2019 2nd Workshop
  on Energy Efficient Machine Learning and Cognitive Computing for Embedded
  Applications (EMC2)}, pp.~21--25, 2019.

\bibitem{Zhang2021fracbnn}
Y.~Zhang, J.~Pan, X.~Liu, H.~Chen, D.~Chen, and Z.~Zhang, ``{FracBNN: Accurate
  and FPGA-efficient binary neural networks with fractional activations},''
  {\em The 2021 ACM/SIGDA International Symposium on Field-Programmable Gate
  Arrays}, 2021.

\bibitem{eldridge2015}
S.~Eldridge, A.~Waterland, M.~Seltzer, J.~Appavoo, and A.~Joshi, ``Towards
  general-purpose neural network computing,'' in {\em 2015 International
  Conference on Parallel Architecture and Compilation, {PACT} 2015, San
  Francisco, CA, USA, October 18-21, 2015}, pp.~99--112, 2015.

\bibitem{liu2020braggnn}
Z.~Liu, H.~Sharma, J.-S. Park, P.~Kenesei, A.~Miceli, J.~Almer, R.~Kettimuthu,
  and I.~Foster, ``Bragg{NN}: Fast x-ray {B}ragg peak analysis using deep
  learning,'' {\em arXiv preprint arXiv:2008.08198}, 2021.

\end{thebibliography}
\clearpage
\section*{Government License}

The submitted manuscript has been created by UChicago Argonne, LLC,
Operator of Argonne National Laboratory (“Argonne”). Argonne, a
U.S. Department of Energy Office of Science laboratory, is operated
under Contract No. DE-AC02-06CH11357. The U.S. Government retains for
itself, and others acting on its behalf, a paid-up nonexclusive,
irrevocable worldwide license in said article to reproduce, prepare
derivative works, distribute copies to the public, and perform
publicly and display publicly, by or on behalf of the Government.  The
Department of Energy will provide public access to these results of
federally sponsored research in accordance with the DOE Public Access
Plan. http://energy.gov/downloads/doe-public-access-plan.
 
\end{document}